# Plasmon modes in double-layer biased bilayer graphene

Nguyen Van Men[1,2], Nguyen Quoc Khanh[2,3], and Dang Khanh Linh[4*]

[1]*An Giang University, 18-Ung Van Khiem Street, Long Xuyen, An Giang, Viet Nam (Email: nvmen@agu.edu.vn).*
[2]*Vietnam National University, Ho Chi Minh City, Viet Nam.*
[3]*University of Science, 227-Nguyen Van Cu Street, 5th District, Ho Chi Minh City, Viet Nam (Email: nqkhanh@hcmus.edu.vn).*
[4] *Ho Chi Minh City University of Education, 280 An Duong Vuong Street, 5th District, Ho Chi Minh City, Vietnam*

**Abstract**

We investigate zero-temperature plasmon modes in a double-layer bilayer graphene structure under a perpendicular electrostatic bias. The numerical results demonstrate that there exist two collective modes which are undamped in the long wavelength limit. The finite potential bias decreases remarkably the plasmon energy in a wide range of wave-vector and makes plasmon branches become Landau damped at a higher wave-vector as compared to unbiased case. We find that the dependence of plasmon dispersions on the system parameters such as the inter-layer separation and carrier density is similar in two cases with and without electrostatic bias.

## 1. Introduction

The unique properties and potential applications of graphene have attracted a large attention of researchers since it was discovered [1-13]. Monolayer graphene (MLG) is a sheet of carbon atoms arranged in a honeycomb planar structure and bilayer graphene (BLG) consists of two MLG layers separated by a small distance $d = 3.35$ Å. The interaction between electrons in two graphene sheets lead to distinct characters of quasi-particles in BLG as chiral massive fermions, compared to chiral massless fermions in MLG and non-chiral massive fermions in ordinary two-dimensional electron gas (2DEG) [14-17]. Recent theoretical and experimental researches demonstrate that an energy gap between conduction and valence bands of BLG can be opened by an electrostatic bias. It has been shown that the characteristics of biased bilayer graphene (BBLG) such as collective excitations, screening and transport properties differ remarkably from those of BLG [18-26].

Collective excitations in layered structures have been intensively studied and applied during the past many years. It was found that collective excitations in graphene have different features, in comparison with those in semiconductor materials [14, 27-32]. Plasmon properties in double bilayer graphene (DBLG) and multilayer bilayer graphene systems have been investigated by many authors for the gapless case [15, 33-34]. The effect of energy gap caused by an electrostatic potential bias between the two graphene layers of BLG on collective excitations at zero temperature has been studied recently by the authors of Ref. [18]. In this paper, we investigate zero-temperature plasmon dispersions in a double layer system consisting of two parallel biased bilayer graphene (hereafter referred as DBBLG) sheets separated by a dielectric spacer as shown in Fig. 1.

## 2. Theory

The plasmon modes can be found from the zeroes of the dynamical dielectric function [27-34],

$$\varepsilon(q, \omega_p - i\gamma) = 0 \qquad (1)$$

where $\omega_p$ is the plasmon frequency at a critical wave vector $q$ and $\gamma$ is the damping rate of plasma oscillations.

---

[*] Corresponding author

*E-mail address:* dangkhanhlinh@tdtu.edu.vn (D.K.Linh).

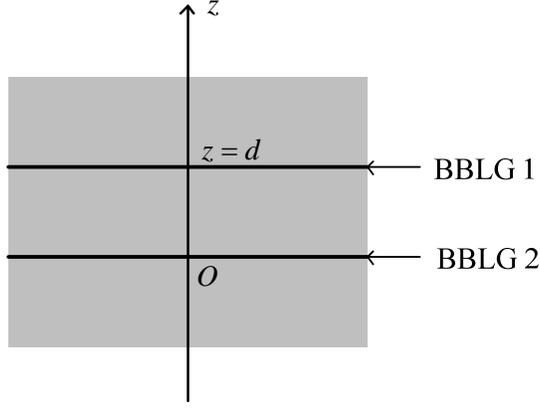

**Fig. 1.** A DBBLG system immersed in a dielectric medium.

In case of weak damping ($\gamma \ll \omega_p$), the plasmon dispersion can be determined by solving the zeros of the real part of the dielectric function [27-34]

$$\mathrm{Re}\,\varepsilon(q,\omega_p) = 0 \tag{2}$$

The dynamical dielectric function of double-layer systems can be written as

$$\varepsilon(q,\omega) = \left[1 - v_{11}(q)\Pi_1(q,\omega)\right]\left[1 - v_{22}(q)\Pi_2(q,\omega)\right] - v_{12}^2(q)\Pi_1(q,\omega)\Pi_2(q,\omega) \tag{3}$$

where $v_{ii}(q)$ and $v_{ij}(q)$ are intra- and interlayer Coulomb bare interactions, respectively [27-30, 34]

$$v_{ii}(q) = \frac{2\pi e^2}{\kappa q}, \tag{4}$$

$$v_{ij}(q) = v_{ii}(q) e^{-qd} \tag{5}$$

and $\Pi_i(q,\omega)$ ($i=1,2$) is the polarization function of $i$-th layer. Within the random-phase-approximation, these functions read [18]

$$\Pi(q,\omega) = g \sum_{\lambda,\lambda',\vec{k}} \left|g_{\vec{k}}^{\lambda,\lambda'}(\vec{q})\right|^2 \frac{f(E_{\vec{k}+\vec{q}}^{\lambda'}) - f(E_{\vec{k}}^{\lambda})}{\omega + E_{\vec{k}+\vec{q}}^{\lambda'} - E_{\vec{k}}^{\lambda} + i\delta}. \tag{6}$$

where the factor $g = 4$ describes the spin and valley degeneracy of the electronic states. Here $\left|g_{\vec{k}}^{\lambda,\lambda'}(\vec{q})\right|^2$ is the vertex factor [18],

$$\left|g_{\vec{k}}^{\lambda,\lambda'}(\vec{q})\right|^2 = \frac{1}{2}\left[1 + \lambda\lambda' \cos\alpha_{\vec{k}} \cos\alpha_{\vec{k}+\vec{q}} + \lambda\lambda' \sin\alpha_{\vec{k}} \sin\alpha_{\vec{k}+\vec{q}} \cos(2\theta_{\vec{k}} - 2\theta_{\vec{k}+\vec{q}})\right] \tag{7}$$

with

$$\tan\alpha_{\vec{k}} = \frac{\hbar^2 k^2}{m^* U}, \tag{8}$$

$$E_{\vec{k}}^{\lambda} = \lambda \sqrt{\left(\frac{U}{2}\right)^2 + \left(\frac{\hbar^2 k^2}{2m^*}\right)^2} \tag{9}$$

is the energy of electrons in BBLG near Dirac points $K$ and $K'$, and $f(x)$ is the Fermi-Dirac distribution function.

Using the equations (3)-(9) we have solved Eq. (2) numerically to obtain the plasmon frequency of the DBBLG system.

## 3. Results and discussions

We assume in the whole paper that the system is immersed in a dielectric medium with an averaged permittivity $\kappa = 2.4$ and two BLG layers have the same carrier density. The Fermi energy and the Fermi wave vector of corresponding unbiased system is denoted by $E_F$ and $k_F$, respectively.

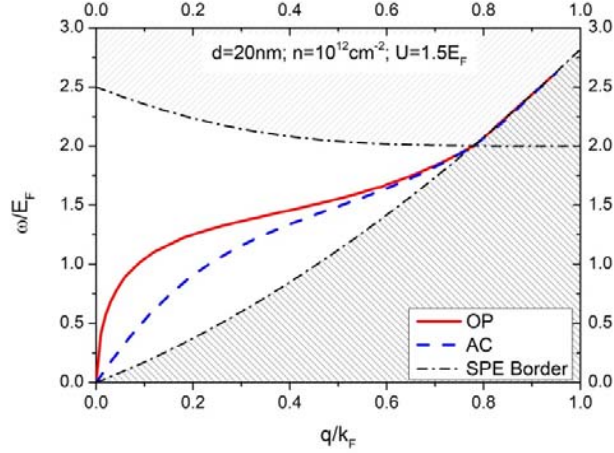

**Fig. 2.** Plasmon modes in DBBLG for $d = 20nm$, $n_1 = n_2 = 10^{12} cm^{-2}$ and $U = 1.5 E_F$. Dashed-dotted lines show the SPE boundaries.

We show in Fig. 2 the plasmon dispersions in DBBLG with $d = 20nm$, $n_1 = n_2 = 10^{12} cm^{-2}$ and $U = 1.5 E_F$. We find two collective modes which are undamped in the long wavelength limit. The higher (lower) mode corresponds to the in-phase optical (OP) (out-of-phase acoustical (AC)) oscillations of carriers in two layers. At potential bias $U = 1.5 E_F \approx 72 meV$, the energy of two plasmon branches differ significantly for small wave-vectors and become identical before entering the SPE continuum at about $q = 0.8 k_F$.

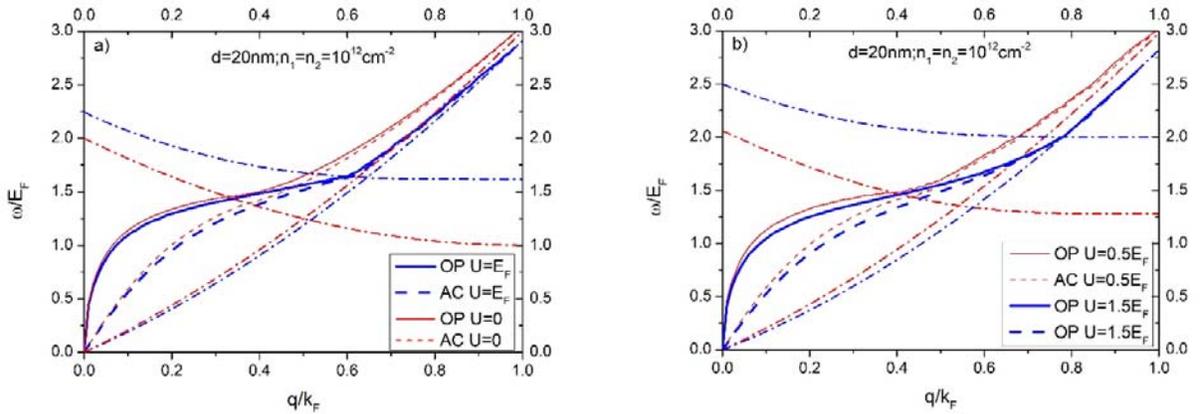

**Fig. 3.** Plasmon frequencies in DBBLG for $d = 20nm$, $n_1 = n_2 = 10^{12} cm^{-2}$ and several values of potential bias. Dashed-dotted lines show the boundaries of the single-particle-excitation (SPE) continuum.

In order to understand the influence of potential bias on plasmon properties of DBBLG, we plot in Fig. 3 plasmon frequencies as a function of wave vector for $d = 20nm$, $n_1 = n_2 = 10^{12} cm^{-2}$ and several values of potential bias. It is seen from the figure that the potential bias, leading to an energy gap in BLG layer, decreases remarkably plasmon frequencies of both branches as in the case of gapped graphene systems [35-40]. In addition, the interband SPE continuum edge increases with increase in the potential bias. Hence both plasmon modes become Landau damped at a higher wave-vectors when the potential bias increases.

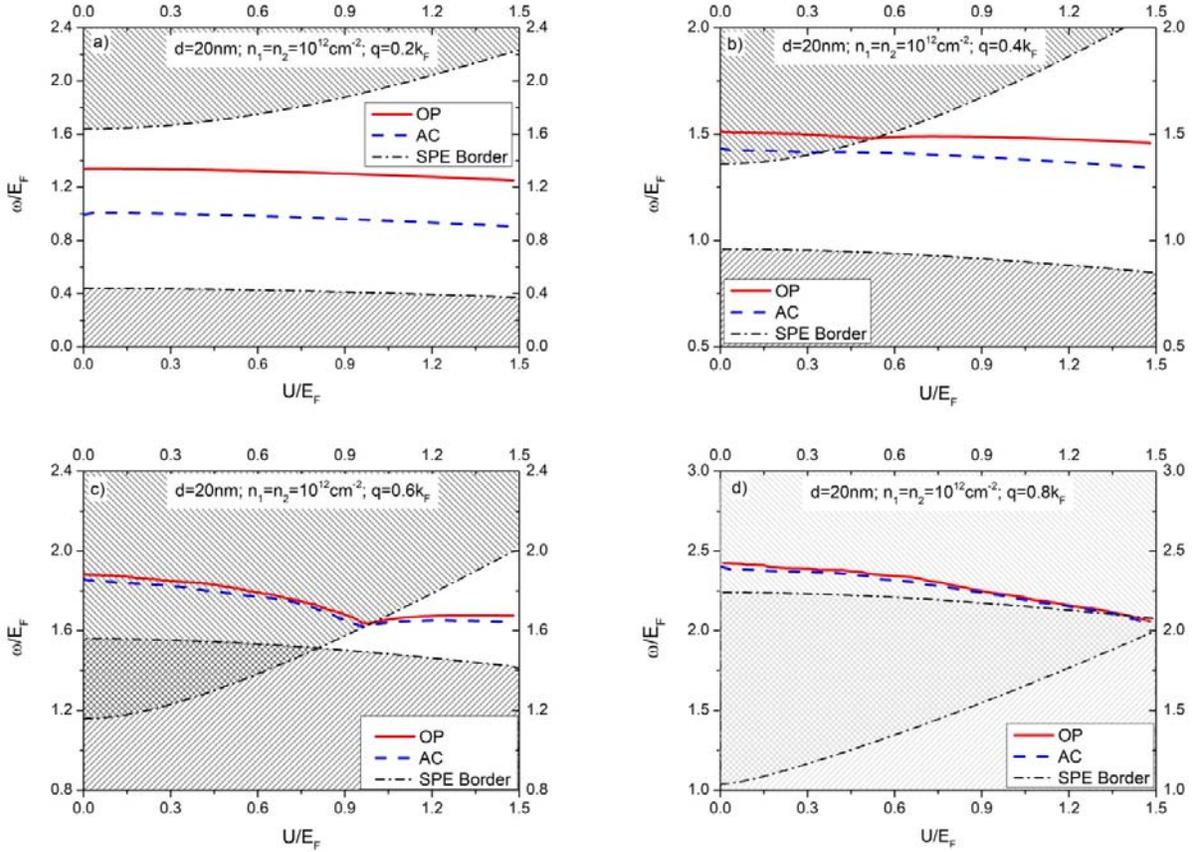

**Fig. 4.** Plasmon frequencies as a functions of potential bias for $q = 0.2k_F$ (a), $q = 0.4k_F$ (b), $q = 0.6k_F$ (c), $q = 0.8k_F$ (d), $d = 20nm$ and $n_1 = n_2 = 10^{12} cm^{-2}$. Dashed-dotted lines show SPE boundaries.

For more information about the effects of potential bias on plasmon characters in DBBLG, we plot in Fig. 4 the frequency of plasmon modes as a function of potential bias varying form zero to $1.5E_F$ for $d = 20nm$, $n_1 = n_2 = 10^{-12} cm^{-2}$ and several values of wave-vectors. From Fig. 4(a) we observe that at $q = 0.2k_F$ the two dispersion curves lie in the SPE continuum gap and both plasmon modes are undamped. For $q \geq 0.4k_F$ the plasmon curves of both branches are in the SPE continuum for small bias and the plasmon modes are Landau damped. When the potential bias increases, the interband SPE continuum edge shifts up and the intraband one shifts down. Hence the dispersion curves can lie in the SPE continuum gap and the plasmon modes become undamped again as shown in Figs. 4(b) and 4(c). From Fig. 4(d) we see that at $q = 0.8k_F$ two plasmon modes merge in the SPE continuum and become damped for all bias values considered.

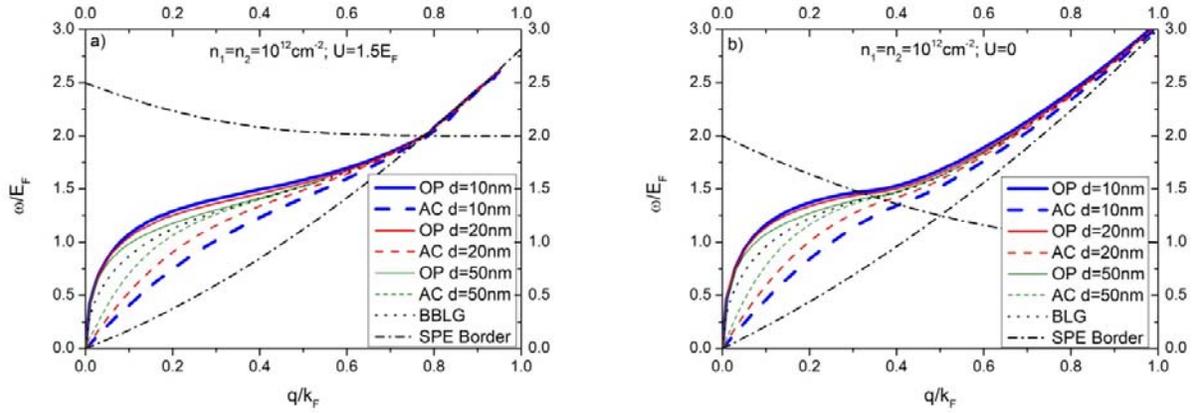

**Fig. 5.** Plasmon dispersions in DBBLG with $U = 1.5E_F$ (a) and in DBLG (b) for $n_1 = n_2 = 10^{12} cm^{-2}$ and several values of the interlayer distance $d$.

We now examine the effects of interlayer separation on plasmon modes by plotting in Fig. 5 the plasmon frequency for $n_1 = n_2 = 10^{12} cm^{-2}$ and several values of the interlayer distance $d$ in two cases $U = 1.5E_F$ (a) and $U = 0$ (b). The figure indicates that the increase in separated distance increases (decreases) pronouncedly AC (OP) plasmon frequency. In the limit $d \to \infty$ two plasmon branches shift more closely to each other and approach gradually to that of single BBLG as in the unbiased case [33-34].

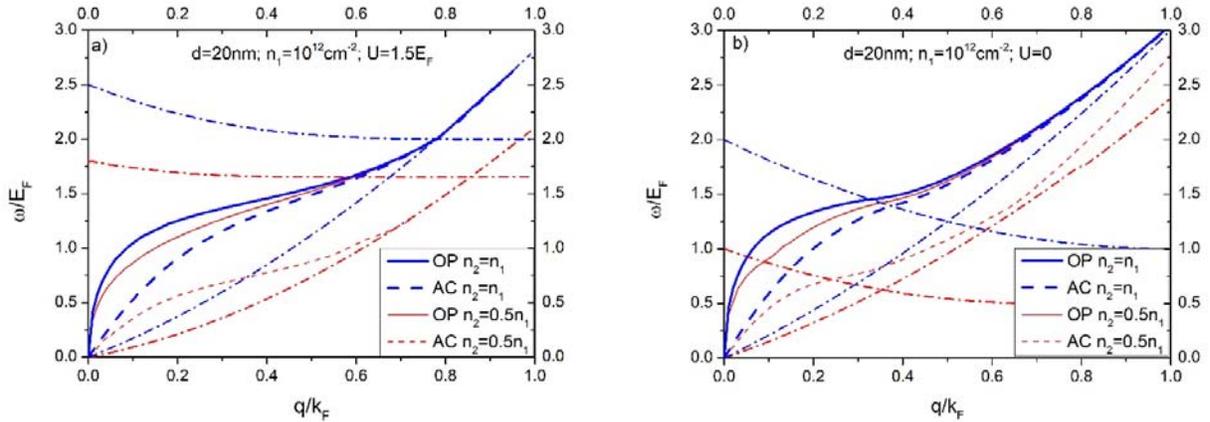

**Fig. 6.** Plasmon modes in DBBLG with $U = 1.5E_F$ (a) and in DBLG (b) for $d = 20nm$ and $n_1 = 10^{12} cm^{-2}$ in two cases $n_2 = n_1$ and $n_2 = 0.5n_1$. Dashed-dotted lines show SPE boundaries.

Finally, to study the effects of the imbalance in carrier density we plot in Fig. 6 the plasmon frequency in DBBLG with $U = 1.5E_F$ (a) and in DBLG (b) for $d = 20nm$ and $n_1 = 10^{12} cm^{-2}$ in two cases $n_2 = n_1$ and $n_2 = 0.5n_1$. As this figure shows, both OP and AC plasmon frequencies decrease with decreasing $n_2$ and the AC branch is affected by the carrier density imbalance more strongly than the OP one is.

## 4. Conclusion

In summary, we have calculated, for the first time, the zero-temperature plasmon dispersions in DBBLG using the random phase approximation. We have found two collective modes which are undamped in the long wavelength limit. The potential bias decreases remarkably plasmon frequencies of both OP and AC branches and with increasing bias the plasmon modes become Landau damped at a higher wave-vectors. We have shown that for both cases with and without bias the increase in separated distance increases (decreases) pronouncedly AC (OP) plasmon frequency and two plasmon modes become identical in the limit $d \to \infty$.